\documentclass[12pt]{article}

\begin{document}
\begin{flushright}
\hfill{SLAC-PUB-7840}\\
\hfill{SU-ITP-98/36}\\
\hfill{hep-th/9805217}\\
\hfill{May 1998}\\
\end{flushright}

\vspace{20pt}

\begin{center}
{\large {\bf Near Horizon Superspace}}

\vspace{30pt}

{\bf Renata Kallosh$^{1a}$, J. Rahmfeld$^{1b}$ and Arvind Rajaraman$^{1
2
\,cd}$}

\vspace{20pt}
$\phantom{a}^1${\it Department of Physics, Stanford University,
Stanford, CA 94305-4060}

\vspace{10pt}
$\phantom{a}^2${\it Stanford Linear Accelerator Center, Stanford
University,
Stanford, CA 94309}

\vspace{80pt}

\underline{ABSTRACT}

\end{center}

The $adS_{p+2}\times S^{d-p-2}$ geometry of the near horizon branes
is promoted to a supergeometry: the solution of the supergravity
constraints for the vielbein, connection and form superfields are
found. This supergeometry can be used for the construction of
new superconformal theories.  We also discuss the Green-Schwarz
action for a type IIB string on $adS_5 \times S_5$.

{\vfill\leftline{}\vfill
\vskip  30pt
\footnoterule
\noindent
{\footnotesize
$\phantom{a}^a$ e-mail: kallosh@physics.stanford.edu. }  \vskip  -5pt

\noindent
{\footnotesize
$\phantom{a}^b$ e-mail: rahmfeld@leland.stanford.edu. }  \vskip  -5pt

\noindent
{\footnotesize
$\phantom{a}^c$ e-mail: arvindra@leland.stanford.edu. }  \vskip  -5pt

\noindent
{\footnotesize
$\phantom{a}^d$  Supported in part by the Department of Energy under
contract No.
DE-AC03-76SF00515.}

\pagebreak
\setcounter{page}{1}

There has been great interest recently in type IIB theory on
$adS_5 \times S_5$ \cite{Maldacena, Gubser,Witten},
due to its possible relation to N=4,d=4
Yang-Mills theory. This vacuum is the near horizon limit of
a D3-brane geometry. Related to this, we also have the
$adS_7 \times S_4$and $adS_4 \times S_7$ vacua of M-theory, which
are the near horizon geometries of a M5-brane and an M2-brane
respectively.

We would like to formulate a description of these vacua in
superspace. This is of great use, since the knowledge of the
supervielbeins  enables us to
write down worldvolume actions of branes in these backgrounds,
in particular, the Green-Schwarz
action for a type IIB string on $adS_5 \times S_5$.

A first step was taken in \cite{KR}, where the torsion and
curvature superfields were found in the framework of the standard
supergravity superspace \cite{CF,BH,HW}. This knowledge enables one to
prove that these vacua are exact solutions. In this paper, we
will find in addition the
vielbeins, connections and form superfields.

In the $adS_5\times S^5$ case, this problem was attacked by
Metsaev and
Tseytlin \cite{Tseytlin} using the  group manifold approach. We shall
extend
their construction to the M-theory vacua, and find closed form
expressions for the supervielbeins and superconnections. This
in particular enables one to directly write down a closed form
expression for the Green-Schwarz action
for a type IIB  on $dS_5 \times S_5$, using \cite{GHMNT}.

We define the forms, following  \cite{Fre} where supergravity
was defined in a group manifold approach:
\begin{eqnarray}
L^a (x, \theta) &=& E^a{} _\mu (x, \theta) dx^\mu +
E^a{}_{\underline \alpha}
(x, \theta) d\theta ^{\underline \alpha}\nonumber \\
L^\alpha (x, \theta) &=& E^\alpha {} _\mu (x, \theta) dx^\mu +
E^\alpha {}_{\underline \alpha} (x, \theta) d\theta ^{\underline
\alpha}\nonumber\\
\omega^{ab} (x, \theta) &=& \omega^{ab} {} _\mu (x, \theta) dx^\mu +
\omega^{ab} {}_{\underline \alpha} (x, \theta) d\theta ^{\underline
\alpha}\label{general}
\end{eqnarray}
where $L^a$ ($L^\alpha$) corresponds to the bosonic(fermionic)
components of the vielbein ($E^a, E^\alpha$ in
\cite{CF}) and $\omega^{ab}$ is the connection
superfield. The lowest ($\theta=0$) components of $E_\mu^a$
and $\omega_\mu^{ab}$ are the space-time vielbein and
connection.

The space-time form-fields
of supergravity $A_{\mu_1, \dots \mu_{p+1}}(x) $ are also introduced in
superspace with the help of the superspace forms. In terms of the
vielbein $L^{\hat A}= (L^a, L^\alpha) $ the form fields are
\begin{equation}
A= L^{\hat A_1} \dots L^{\hat A_{p+1}} A_{\hat A_1 \dots \hat A_{p+1}}
(x,
\theta),\qquad
F=dA,
\qquad
dF=0
\end{equation}
In the group-manifold approach, the curvature and torsions are
defined somewhat differently. In effect, non-zero value of the
curvature and torsion are absorbed by redefining them in such
a way that the new curvature and torsion are zero.
For example, the standard definition of $adS$ geometry has a
non-zero curvature
\begin{equation}
R_{ab}{}^{cd}=-  \delta _{[a}{}^c \delta _{b]}{}^d\end{equation}
which translates in form language to
\begin{equation}
R^{cd} \equiv (d \omega + \omega \wedge \omega)^{cd} =- L^c \wedge
L^d.
\label{curv}\end{equation}
Instead, in the group manifold approach one uses the
anti-de-Sitter algebra
\begin{equation}
[P_a,P_b]=J_{ab}\,,
\qquad[P_a,J_{bc}]=\eta_{ab}P_c-\eta_{ac}P_b\,,
\qquad
[J_{ab},J_{cd}]=\eta_{bc}J_{ad}+3\hbox{ terms}\,
\end{equation}
and introduces the 1-form differential operator
\begin{equation}
{\cal D} = d + \omega^{ab} J_{ab} + L^a P_a.
\end{equation}
The $adS$ curvatures are now defined by
\begin{equation}
{\cal D}^2 \equiv {\cal R}^{ab} J_{ab} + {\cal R}^a P_a
\label{D2}\end{equation}
and it is easily verified that
\begin{eqnarray}
{\cal R} ^{ab} &=&  R^{ab}  + L^a
\wedge L^b \\
{\cal R} ^{a} &=& T^a
\end{eqnarray}
which vanish  for this geometry. Hence, the $adS$ geometry
can be defined by requiring vanishing  generalized
$adS$ curvatures.

For the analogous description in superspace we must generalize
the differential operator ${\cal D}$ to include all the generators of
the
appropriate superconformal algebra:
$OSp(8|4)$
for the $adS_{4}\times S^{7}$ vacuum of M theory, $OSp(6,2|4)$ for the
$adS_{7}\times S^{4}$ vacuum of M theory and $SU(2,2|4)$
for the $adS_{5}\times S^{5}$
vacuum of IIB string theory.
In general, the superconformal algebra is of the form
\begin{eqnarray}
\left [B_A ,  B_B \right ] &=& f_{AB}^C B_C \nonumber \\
\left [F_{\alpha} ,   B_B \right ]  &=&  f_{\alpha B}^\gamma F_\gamma
\nonumber\\
\left\{ F_\alpha  , F_\beta \right\}&=&  f_{\alpha \beta}^C B_C
\label{algebra}\end{eqnarray}
where $B_A$ and $F_\alpha$ are the bosonic and fermionic
generators, and the $f$ are the structure constants.
The differential operator is then
\begin{equation}
{\cal D}=d+L^A B_A+L^\alpha F_\alpha.
\end{equation}
$L^A$ and $L^\alpha$ are the left-invariant Cartan forms. The
equation ${\cal
D}^2=0$ leads then to the
Maurer-Cartan equations
\begin{eqnarray}
dL^A&+&L^B \wedge L^C f_{BC}^A -L^\alpha \wedge L^\beta f_{\alpha
        \beta}^A=0 \\
dL^\alpha&+&L^A \wedge L^\beta f_{A\beta}^\alpha =0.
\end{eqnarray}

We will now solve these equations using the standard method which
was also employed in \cite{Tseytlin}.
Let $G$ be an element of the superconformal group, then
\begin{equation}
 G^{-1}d G
  =L^A B_A +L^\alpha F_\alpha.
\end{equation}
Writing $G=g(x)e^{\theta F}$, we find that
\begin{equation}
  G^{-1}d G=e^{-\theta  F}  D  e^{\theta F}, \label{prop}
\end{equation}
where
 \begin{equation}
D =d+ L^A_0 B_A
\end{equation}
 with $L^A_0= L^A(x,\theta=0)$.
It is now useful to decompose $L$ as
\begin{equation}
L=L_0(x)+\tilde L(x,\theta). \label{split}
\end{equation}
To find explicitly $\tilde L^A$ and
$\tilde L^\alpha$ one
then replaces $\theta\rightarrow t \theta$ and introduces
a $t$-dependence into the vielbein components (which
will be eventually be removed by setting $t=1$), giving rise
to the modified relation
\begin{equation}
e^{-t\theta F}  d e^{t \theta F}=
\tilde L_t^A B_A +\tilde L_t^\alpha F_\alpha.
\end{equation}
Differentiating both sides and utilizing the superalgebra
(\ref{algebra})
one finds  the differential equations:
\begin{eqnarray}
\partial_t \tilde L_t^A & = & \theta^\alpha f_{\alpha \beta}^A
     \tilde L_t^\beta
   \\
\partial_t \tilde L_t^\alpha & = & D \theta^\alpha
         -\theta^\gamma f_{\gamma A}^\alpha \tilde L_t^A.
\end{eqnarray}
which have the structure of coupled harmonic oscillators.
Together with the initial conditions
\begin{equation}
\tilde L_{t=0}^A=\tilde L_{t=0}^\alpha=0
\end{equation}
and (\ref{split}), we can easily write down  the
explicit solution for the supervielbein  (setting $t=1$)
in closed form.
One finds that
\begin{equation}
L^\alpha =\left({\sinh {\cal M}  \over {\cal
M}}\right)^\alpha_\beta
(D\theta)^\beta
\label{WZ1} \end{equation}
and
\begin{equation}
L^A =L^A_0+2 \theta^\alpha  f_{\alpha \beta}^A
\left({\sinh^2 {\cal M}/2  \over {\cal
M}^2}\right)
^\beta_\gamma
(D\theta)^\gamma,
\label{WZ2}\end{equation}
where
\begin{equation}
\left({\cal M}^2\right)^\alpha_\beta =-\theta^\gamma f_{\gamma
A}^\alpha\theta^\delta f_{\delta \beta}^A.
\label{2d}\end{equation}
Note that ${\cal M}^2$ is quadratic in $\theta$ and that all higher
order terms
in $\theta$ in eqs. (\ref{WZ1}), (\ref{WZ2}) are given by even
powers of ${\cal
M}$, i.e. by powers of ${\cal M}^2$ up to  $({\cal M}^2)^{16}$.

If we choose $\theta$ to be the standard fermionic coordinates
of superspace, the above solution gives the superspace geometry
in the Wess-Zumino gauge. This can be readily verified by noting
that
\begin{equation}
(D \theta)^\alpha= d\theta^\alpha +( L^A_0 B_A  \theta) ^\alpha,
\end{equation}
hence $L^{\alpha}_{\underline\beta}(x,\theta=0)=
\delta^\alpha_{\underline\beta}$.

However, it turns out that there is an often more
convenient gauge which simplifies the geometry
considerably. Consider space-time dependent $\theta^\alpha$
of the form
\begin{equation}
\theta^{\alpha}(x) = e^{\alpha} _{\underline {\alpha}}(x)
\theta^{\underline
{\alpha}}
\end{equation}
so that
\begin{equation}
(D\theta)^\alpha =e^\alpha_{\underline \alpha}(x) d \theta ^{\underline
\alpha}
+\left (  (d+ L^A_0 B_A)^{\alpha}_ \beta  e^\beta _{\underline
\alpha}(x) \right
)\theta ^{\underline \alpha}
\end{equation}
The second term drops if we choose $\theta$ to
be the Killing spinors of the background! As is well-known,
those are precisely defined by the equation
\begin{equation}
 (d+ L^A_0 B_A)^{\alpha}_ \beta  \epsilon^\beta  (x)_{Kill} =0
\end{equation}
and the solution  is of the form \cite{LPR}
\begin{equation}
\epsilon^{\alpha} (x)_{Kill} = e^{\alpha} _{\underline {\alpha}}(x)
\epsilon_{\rm const}^{\underline {\alpha}}.
\end{equation}
where $\epsilon_{\rm const}^{\underline {\alpha}}$ is a constant spinor.
Hence, choosing $\theta=\epsilon_{Kill}(x)$ in (\ref{prop}) leads to
\begin{equation}
\qquad
(D\theta)^\alpha =e^\alpha _{\underline \alpha}(x) d \theta ^{\underline
\alpha}.
\end{equation}
With the explicit Killing spinors, constructed for all relevant
$adS_*\times S^*$ spaces in \cite{LPR},
the superspace structure simplifies as follows:

\begin{equation}
L^\alpha =e^\alpha_{\underline \alpha} (x, \theta) d \theta ^{\underline
\alpha} \qquad L^A =e^A_M (x, \theta=0) dx^M  + e^A_{\underline
\alpha}(x,
\theta)
  d \theta ^{\underline \alpha},
\label{super1}\end{equation}
where
\begin{equation}
e^\alpha_{\underline \alpha} (x, \theta) =\left({\sinh {\cal M}
\over {\cal
M}}\right)^\alpha_\beta
 e^\beta_{\underline
\alpha}(x) \ ,
\end{equation}
and
\begin{equation}
e^A_{\underline \alpha}(x, \theta)
=  \theta^\alpha  (x)  f_{\alpha
\beta}^A
\left({\sinh^2 {\cal M}/2  \over {\cal
M}^2}\right)
^\beta_\gamma
e^\gamma_{\underline \alpha}(x)
\end{equation}
where ${\cal M}$ now contains the space-time dependent spinors
$\theta^{\alpha}(x)$
\begin{equation}
({\cal M}^2)^\alpha_\beta= -\theta^\delta (x)  f^\alpha _{\delta
A}\theta^\gamma
(x)   f^A _{\gamma \beta}.
\label{M2}\end{equation}
Instead of the standard Wess-Zumino gauge with
$e^\alpha_{\underline \alpha} (x, \theta=0)=
\delta^\alpha_{\underline \alpha}$
we have a Killing spinor gauge
since the fermion-fermion part of the vielbein at $\theta=0$ is
`curved'  with
\begin{equation}
E^\alpha_{\underline \alpha} (x, \theta=0)=e^\alpha_{\underline \alpha}(x).
\end{equation}
Like in the flat superspace, there are no fermions, the gravitino
vielbein component $ \psi^\alpha {} _\mu (x, \theta) dx^\mu$ is not
generated
at all levels of $\theta$. This also explains why the boson-boson
component of
the vielbein $e^A_M (x, \theta=0)$ is $\theta$-independent.

After developing the general formalism let us now turn to a
specific example, the type $IIB$ theory on $adS_{5}\times S^{5}$.
This vacuum is a special case of the IIB
superspace \cite{HW}. It was presented in
\cite{KR} in eqs. (46-51). The  expressions for
the constant Lorentz curvature and torsion of
the string
vacuum can be
reinterpreted via the superconformal group manifold approach. Following
\cite{Tseytlin} we consider a coset superspace $SU(2,2|4) \over
SO(4,1)\times
SO(5)$
with the even part
$adS_5\times S^5 = {SO(4,2) \over SO(4,1)} \times
{SO(6) \over SO(5)}$.
The   even generators are then two pairs of
translations and rotations,
 $(P_a, J_{ab})$ with  $a=0,1,2,3,4$ for $adS_5$ and
$(P_{a'},J_{a'b'}), a'=
5,6,7,8,9$
 for $S^5$  and the odd generators are the two
 $D=10$ Majorana-Weyl spinors
 $Q_{\alpha \alpha' I}$. The differential operator related to
$SU(2,2|4)$
superalgebra is given by
\begin{equation}
{\cal D} =d +  G^{-1} dG  = d +  L^{a} P_{a}+L^{ab} J_{ab}+
                L^{a} P_{a}+L^{ab} J_{ab}+L^{\alpha\alpha'I}
                Q_{\alpha\alpha'I}
\end{equation}
satisfying
$ {\cal D}^2  =0.
$
Taking care of the change of
notations with respect to the different representation for the
fermions and
splitting the ten-dimensional $\hat a$ into two five-dimensional
ones $a, a'$
one can verify that the supergravity constraint for the vector part of
the
torsion
can be
brought to the form
\begin{eqnarray}
{\cal R}^{a}&=&  dL^{a} + L^b \wedge L^{ba} - i \bar L^I \gamma^a
\wedge L^I
=0\nonumber\\
{\cal R}^{a'} &=&  dL^{a'} + L^{b'} \wedge L^{b'a'} - i \bar L^I
\gamma^{a'}
\wedge L^I  =0
\end{eqnarray}
where we follow the conventions of \cite{Tseytlin}.
For the spinorial part of the
torsion one finds
the following form (upon changing the spinorial representation)
\begin{eqnarray}
{\cal R}^{\alpha \alpha' I}&=&  dL^{\alpha \alpha' I} +({i\over 2}
\gamma^a
\epsilon^{IJ} L^J \wedge L^a -{1\over 4} \gamma^{ab}  L^I \wedge
L^{ab})^{\alpha \alpha'} + \dots =0
\end{eqnarray}
The Lorentz curvature form of supergravity has a non-vanishing
fermion-fermion
component as well as
the
boson-boson component. It can easily be shown that these equations
are equivalent to the constraints of the $adS_5\times S^5$
discussed in \cite{KR}.
Thus the results for the Lorentz-valued curvatures and torsions can be
rewritten as
${\cal D}^2=0$ where $\cal D$ is based on superconformal algebra. In
addition,
the supergravity  $adS_{5}\times S^{5}$ vacuum is defined by the
presence of the 3-form and 5-form.
\begin{eqnarray}
F_{(3)}&=& -i L^{\hat a}  \wedge  L^{\gamma}( \sigma^{\hat a})
_{\gamma\delta}
\wedge
L^{\delta} + c.c.\\
G_{(5)}&=& L^{ a}\wedge L^{ b}\wedge L^{c } \wedge L^{d} \wedge L^{e}
\epsilon_{abcde}+  L^{ a'}\wedge L^{ b'}\wedge L^{c' } \wedge L^{d'}
\wedge L^{e'} \epsilon_{a'b'c'd'e'}\nonumber\\
&+& L^{\hat a}\wedge L^{\hat b} \wedge L^{\hat c } \wedge L^{\gamma}
(\sigma_{\hat a \hat b\hat c})_{ \gamma\delta}\wedge L^{\delta}
\end{eqnarray}
Here we use  the notation of  \cite{HW} but for convenience we  split the
ten-dimensional vector index $\hat a = a, a'$.

We now turn to the supergeometry of this space following
our general procedure and using  the conventions of
\cite{Tseytlin}.
The complete superspace forms in the Killing gauge  specified for
the
$adS_{5} \times S^{5}$ are
\begin{equation}
L^{\alpha \alpha' I}=e^{\alpha \alpha' I} _{\underline \alpha \underline
{\alpha'} \underline I } (x, \theta) d \theta ^{\underline \alpha
\underline
{\alpha'} \underline I }
\end{equation}
\begin{eqnarray}
L^{\hat a }&=&e^{\hat a }_m (x, \theta=0) dx^m  + e^{\hat a
}_{\underline
\alpha \underline {\alpha'} \underline I }(x, \theta)
  d \theta ^{\underline \alpha \underline {\alpha'} \underline I } \\
L^{\hat a \hat b} &=&\omega^{\hat a \hat b} _m (x, \theta=0) dx^m  +
e^{\hat a
\hat b} _{\underline \alpha \underline {\alpha'} \underline I }(x,
\theta)
  d \theta ^{\underline \alpha \underline {\alpha'} \underline I }
\end{eqnarray}
\

Here
\begin{equation}
e^{\alpha \alpha' I} _{\underline \alpha \underline {\alpha'}
\underline I }
(x, \theta)  =\left({\sinh {\cal M}  \over {\cal
M}}\right)^{\alpha \alpha' I} _{\beta \beta' J} e^{\beta \beta' J}
_{\underline
\alpha \underline {\alpha'} \underline I } (x, \theta=0)  \ ,
\end{equation}

\

\begin{equation}
e^{\hat a} _{\underline \alpha \underline {\alpha'} \underline I
}(x, \theta)
= -2i ( \bar \theta^I (x)\gamma^{\hat a})_{\alpha \alpha'}
\left({\sinh^2
{\cal M}/2  \over {\cal
M}^2}\right)
^{\alpha \alpha'
I} _{\beta
\beta' J} e^{\beta \beta' J} _{\underline \alpha \underline {\alpha'}
\underline I }(x, \theta=0)
\end{equation}

\begin{equation}
e^{\hat a b} _{\underline \alpha \underline {\alpha'} \underline I
}(x, \theta)
= 2 \epsilon ^{IJ} ( \bar \theta^I (x)\gamma^{\hat a \hat
b})_{\alpha  \alpha'}
  \left({\sinh^2 {\cal M}/2  \over {\cal
M}^2}\right)
^{\alpha
\alpha' J} _{\beta \beta' K} e^{\beta \beta' K} _{\underline \alpha
\underline
{\alpha'} \underline I }(x, \theta=0)
\end{equation}
where
\begin{eqnarray}
({\cal M}^2)^{ I} _{ L}&=& [ \epsilon^
{IJ}
(-\gamma^{ a}  \theta^{J}(x) \bar \theta^L(x) \gamma^{ a} + \gamma^{ a'}
\theta^{J}(x) \bar \theta^L(x) \gamma^{ a'} )\nonumber\\
&+& {1\over 2}
\epsilon^{KL} (\gamma^{ab} \theta^I (x)\bar \theta^K (x)\gamma^{ab}
+\gamma^{a'b'} \theta^I(x) \bar \theta^K(x) \gamma^{a'b'})
] \label{msquare}
\end{eqnarray}
We use the simplifying notation $\theta^{\alpha \alpha' I} (x) = \theta
^{\underline \alpha \underline {\alpha'} \underline I } e^{\alpha
\alpha' I}
_{\underline \alpha \underline {\alpha'} \underline I } (x) $. We
expect that
this form of the superspace will be most suitable for the
construction of the
superconformal D3 brane action.

It may also be useful to work within the
Wess-Zumino gauge
in the superspace. The solution is than given by

\begin{equation}
L^{ I}  =\left[ \left({\sinh {\cal M} \over {\cal M}}\right) D\theta \right]^{I}
\label{I}
\end{equation}
and
\begin{eqnarray}
L^{\hat a }&=&e^{\hat a }_m (x) dx^m   -4 i  \bar \theta^I \gamma^{\hat
a}
\left({
\sinh^2  {\cal M}/2 \over {\cal M}^2} D\theta \right)^I\label{La}
\end{eqnarray}
where ${\cal M}^2$ is given by eq. (\ref{msquare}) with constant $\theta$.
Here,
\begin{equation}
(D\theta)^I  =\left ( d  +{1\over 4}(  \omega^{ab} \gamma_{ab} +
\omega^{a'b'}
\gamma_{a'b'}) \right )\theta^I  -{1\over 2} i\epsilon^{IJ} (e^a
\gamma_a +
ie^{a'} \gamma_{a'}) \theta ^J
\end{equation}

The complete form of the GS type IIB  $\kappa$-symmetric string
action  in
the generic supergravity background was presented in \cite{GHMNT}:
\begin{eqnarray}
S =-\frac{1}{2}\int_{\partial {M_3}}  d^2\sigma\  \sqrt{g} \, g^{ij}
 L_i^{\hat a} L_j^{\hat a} +  {\rm i}\int_{M_3}
 s^{IJ}   L^{\hat a} \wedge  \bar{L}^I\gamma^{\hat a}\wedge  L^J
\ ,
\label{action}
\end{eqnarray}
where $S^{IJ}$ has non-vanishing elements $S^{11}= -S^{22}=1$. This
action   in
$adS_{5}\times S^{5}$  background was given in \cite{Tseytlin} up to
terms of
the
order $\theta^4$. The complete action in this background is the one in
eq.
(\ref{action}) where  the pull-back of the space-time forms to the
world-sheet
$L_i^{\hat a} , L_i^{I}$ is obtained from    eqs. (\ref{I}),
(\ref{La}) with $d
\theta \equiv d\sigma^i  \partial_i \theta$ and  $d x \equiv d\sigma^i
\partial_i x$.

In conclusion, we have solved the supergravity superspace constraints
for
$adS_{p+2}\times S^{d-p-2}$ backgrounds.
We have found here the complete expressions for the superfield
values of the
vielbeins, connections and forms in two different gauges.
One gauge that utilizes the space-time Killing spinors of the
background leads to a vanishing gravitino superfield
$E^\alpha_\mu (x, \theta)=0$, and a curved fermion-fermion
vielbein. In this gauge we observe a dramatic simplification
of the superspace structure. The constraints can also be solved
in the standard Wess-Zumino gauge. Here, the gravitino superfield
picks up terms of linear and higher order in $\theta$,
and the fermion-fermion vielbein is flat at $\theta=0$.
For various brane actions both of these gauge choices
may be useful.

Thus we have made here  a necessary step for
constructing the brane actions in $adS_{p+2}\times S^{d-p-2}$
backgrounds. As explained in \cite{conformal} the
classical actions of M2, M5, D3 and some other branes in their
near horizon superspace geometry will have two types of symmetries:
global
superconformal symmetry
due to the isometry of the background, and local reparametrization and
$\kappa$-symmetry. Upon gauge-fixing, the local symmetries of the
gauge-fixed
theory will have some form of non-linearly realized superconformal
symmetry. It
will be a challenge to construct such superconformal actions as well
as to find
the gauges in which the actions take the simplest form.

We have also made progress towards understanding of the GS type IIB
string theory in $adS_{5}\times S^{5}$ background. We extended the
results of \cite{Tseytlin}, where the action was explicitly given up
to $\theta^4$ terms, by solving the constraints exactly which
enabled us to give a complete and closed form of the string action
in this background.

\vskip 2 cm
We had  stimulating discussion  with   P. Howe, J. Kumar, J. Schwarz, E.
Sezgin, D. Sorokin, M. Tonin, P. Townsend, A. Tseytlin, A. Van
Proeyen and P. West.
The work of R.K and J.R is supported by the NSF grant PHY-9219345.
The work of A. R. is also
supported in part by the Department of Energy under contract No.
DE-AC03-76SF00515.


\end{document}